# Engineering Tagging Languages for DSLs

Timo Greifenberg, Markus Look, Sebastian Roidl, Bernhard Rumpe
Software Engineering
RWTH Aachen University
http://www.se-rwth.de/

*Abstract*—To keep a DSL clean, readable and reusable in different contexts, it is useful to define a separate tagging language. A tag model logically adds information to the tagged DSL model while technically keeping the artifacts separated. Using a generic tagging language leads to promiscuous tag models, whereas defining a target DSL-specific tag language has a high initial overhead. This paper presents a systematic approach to define a DSL-specific tag language and a corresponding schema language, combining the advantages of both worlds: (a) the tag language specifically fits to the DSL, (b) the artifacts are kept separated and enabling reuse with different tag decorations, (c) the tag language follows a defined type schema, and (d) systematic derivation considerably reduces the effort necessary to implement the tag language. An example shows that it can at least partially be realized by a generator and applied for any kind of DSL. *Index Terms*—Software Engineering, Modeling, MDE, GSE

## I. Introduction

Model-Driven Software Engineering (MDE) [1] makes use of models written in domain specific languages (DSL) as primary development artifacts. It tries to abstract from technical configurations in order to focus on the application domain itself. Generative Programming [2] uses those models as input for code generators which in turn are able to generate most parts of the software system. Nevertheless, some parts of the software system cannot be generated from only the domain model. Further adaptations of the generated system might be needed which can be done by integration of handwritten code [3] or by using configurable generators [4]. For configurable generators, the domain model on its own is often not sufficient as input. Additionally, further data, e.g. platform specific information needs to be provided.

There are mainly two approaches to provide this additional data: the domain model is enriched by additional information or the additional information is stored in separated artifacts. While the former has the drawback of polluting the domain model by mixing domain information with additional data, the latter has the drawback that different artifacts need to be synchronized over time due to model evolution. Model pollution is especially bad when models are supposed to be reused within several projects using different generators. Moreover, if different technical aspects must be covered, a lot of additional information is needed, which easily raises the degree of model pollution.

Nevertheless a DSL cannot incorporate all additional information, since it is typically outside of the domain but required for code generation. To keep a DSL clean, readable and reusable in different contexts, it is useful to define a separate tagging language. The drawback of the second case may be overcome by automated transformations which are used and successfully applied [5] in Model Driven Architecture (MDA) [6]. But still each kind of additional information has to be considered on the model level, resulting in huge additional effort. In an agile development process, models are often reused and information may change frequently, further increasing the cost of adopting the transformations.

We present our approach for adding additional information to existing models via a separate tagging language without polluting the models and without neccessary model transformations. A tag model logically adds information to the tagged DSL model, respectively its model elements, while technically keeping the artifacts separated. Using a generic tagging language leads to promiscuous tag models, whereas defining a target DSL-specific tag language has a high initial overhead. To reduce this overhead and to avoid promiscuous models, we present a systematic approach to define a DSL-specific tag language and a corresponding schema language, combining the advantages of both worlds. This pair is called tagging languages.

We formalize the specification of additional information and explicitly model it by using tagging languages fitted specifically to the existing modeling language. Additional data is defined within external tag models, preventing model pollution. Moreover, data from different domains like persistency configuration, security configuration or GUI style configuration can be defined in separate tag models. In this way the corresponding domain experts are given a model-based possibility to attach additional information and can work independently from each other given only the information necessary for their work, thus focussing completely on their domain. The situation is even further improved by the second DSL, the tagschema language, allowing for the definition of tag types for a given tag language leading to a type system for tags. Thus, the domain expert, modelling the additional information can be efficiently supported by tools utilizing tagschema models to validate corresponding tag models. Additionally the generator developer is given the possibility to define the kinds of additional information the generator is able to handle. This is similar to the dynamic features pattern presented in [7]. By providing the possibility to separate additional information from each other and from the domain model, we are able to follow the concept of separation of concerns [8]. Consistency problems can be avoided as the specification of additional information takes place at the model level. Thus, tooling can be used to ensure the consistency between models of all involved languages. As both of the mentioned DSLs somehow depend on the existing modeling language, a methodology for creating the corresponding tagging languages is provided. This supports the language developer since we assume that by such rules, the specific language can be created faster and with better quality. Therefore the derivation rules have to be kept as easy as possible. The methodology can be implemented as a derivation algorithm and generated automatically.

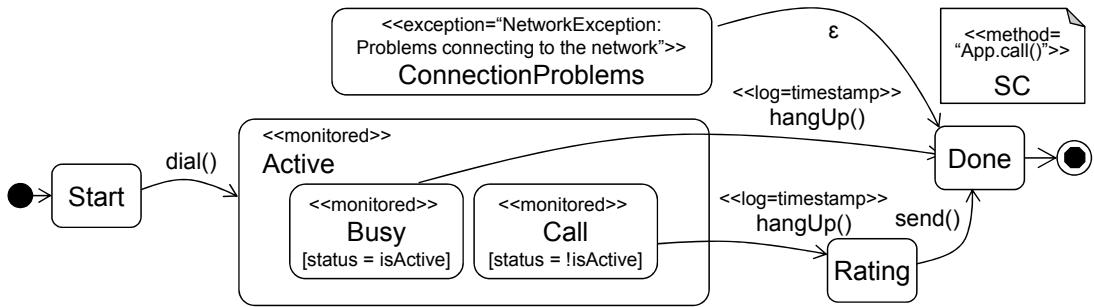

Fig. 1: A graphical representation of a Statechart that models a method of a Voice over Internet (VoiT) App.

The paper is organized as follows. In Section II an example is introduced. In Section III used techniques and technologies are introduced. The overall approach is presented in Section IV. Here, common language parts, derivation rules for the source language specific language parts and necessary context conditions are explained. The derivation process is applied exemplarily in Section V. Related work is presented in Section VI before the paper closes with the conclusion in Section VII.

## II. Tagging Statecharts

To illustrate our methodology, we start out with an example where we tag a Statechart (SC) with additional information. We use Statecharts as predefined in our language family UML/P [4], [9], [10]. The UML/P consists of structural as well as behavioural languages, such as Classdiagramms (CD), Objectdiagrams (OD), Statecharts (SC), Sequencediagrams (SD) and a test case language (TC). The UML/P is close to the UML [11] containing extensions especially suited for being used in a generative software engineering context. Most of these extensions consider the use of the models in combination with a code generator. Thus, concepts such as underspecification, resolving non-determinism or flattening of SCs are directly incorporated in the language family. Furthermore, the syntax of the UML/P is textual, whereas the UML provides a graphical syntax. Nevertheless, we use the graphical UML notation also for visualising UML/P diagrams.

Figure 1 shows a graphical representation of a Statechart that models a method of a Voice over Internet (VoiT) App. When the user wants to make a call, the application starts dialing. If there is currently a call, active the current call cannot be established because the phone is busy. If the called person is not busy, the call is established. After finishing the established call, the user gets the possibility to rate the quality of the call. The rating is sent and the overall method terminates. After a non established call, the method terminates directly. In case there are any problems with the current network connection, the method throws an exception. The modeled SC consists of states and transitions as top level elements. States have a name, may contain substates to support hierarchy and invariants. Transitions connect a source state with a target state. Furthermore, actions are modeled. Figure 1 does not contain all elements but a subset necessary to understand our methodology. A complete overview of all SC elements is provided in [4], [9], [10]. Every element, even the SC itself, may also be extended by stereotypes.

The stereotype «monitored» is added to the `Active` state and its substates, signalling that certain monitoring functions should be available in a respective implementation. The stereotype «log=timestamp» defines that every time the transition is active, a timestamp is written to the log. The state `ConnectionProblems` serves as an exception state. Every time an exception occurs, this state becomes the active state of the SC. This is implicitly expressed by the stereotype `exception`, specifying that there is an implicit transition from every state to the exception state. The value of the stereotype specifies the exception type and a message. Due to the limited expressiveness, this is modeled via a single String separated by a colon. The definition of the SC itself is enriched by a stereotype, as shown in Figure 1, modeling that the SC is used to specify the behaviour of the method `App.call()`. Although the SC is quite small, it shows the problems of using stereotypes in the modeling process: The technology agnostic part of the model is polluted with technology specific information and thus one of the key strengths of model driven development, getting an

```
1  package mobile;
2  statechart Mobile {
3
4    initial state Start;
5
6    state Active {
7      state Call{
8        [status=!isActive];
9      }
10     state Busy {
11       [status=isActive];
12     }
13   }
14
15   state ConnectionProblems;
16
17   final state Done;
18
19   Start -> Active : dial() ;
20   ...
21 }
```

Listing 1: Textual UML/P notation of the SC shown in Figure 1.

```
1  package mobile;
2  conforms to loggingschema.StatechartTagSchema;
3
4  tags StatechartTags for Mobile {
5
6    tag Mobile with Method = "App.call()";
7
8    within Active {
9      tag Call,Busy with Monitored;
10   }
11
12   tag Active with Monitored;
13
14   tag ConnectionProblems with
15     Exception {
16     type = "NetworkException",
17     msg = "Problems connecting to the mobile network!";
18       };
19   ...
20 }
```

Listing 2: A valid tag model, written in a textual language enriching the SC shown in Listing 1

```
1  package loggingschema;
2  tagschema StatechartTagSchema {
3
4    tagtype Monitored for State;
5
6    tagtype Log:["timestamp"|"callerID"] for Transition;
7
8    tagtype Method:String for Statechart;
9
10   tagtype Exception for State {
11     type:String,
12     msg:String;
13   }
14 }
```

Listing 3: A tagschema, written in a textual language defining the tag types used in Listing 2.

easy overview of the domain, is lost. Obviously, this gets worse when more additional information is required, the modelled system is more complex or when multiple roles or concerns are involved. To overcome this problem, the additional information is externalized into a separate tag model. Listing 1 conveys the same information as Figure 1 in the textual syntax of the UML/P without additional information, which is shown in Listing 2 as an external, textual tag model.

Listing 2 shows a valid tag model, written in a textual language derived using our proposed methodology. Like the SCs, each tag model starts with a package declaration that establishes a namespace, followed by the keyword `conforms to` and a reference to a tagschema, shown in Listing 3. Note that the tagschema also contains a package delcaration that establishes a namespace and is used to fully qualify the name of the tagschema. By using the keyword `conforms to`, each tag file declares conformity to one or more tagschemas, as shown in Listing 2 line 2, expressing that tags used within the tag model are defined within the tag schema model. We assume that different technological specifica for a model will lead to different sets of tags, which are free of conflict by definition. Conflicting tags might however be possible but are not handled yet. A formal specification of this relation is given in Section IV, where we also explain how the name references between the different models are resolved. In the example, the two substates `Busy` and `Call` of the `Active` state are tagged with `Monitored` by navigating into it. Listing 2 contains the same information, that was modeled as stereotypes in Figure 1. It should be noted, that we consider four different tag types, which are explained in detail in Section IV-B. Listing 3 shows the schema for the presented tag model, which also starts with a package declaration. The schema contains the four different tag types: `Monitored` that only consists of a single word as key, `Log` which may take two disjoint values, `Method` that can take an arbitrary String as value and `Exception` that has multiple fields as values. All these tag types are defined in Listing 3.

## III. MONTICORE

Our methodology to derive the tagging languages is based on the modular language workbench MontiCore [12], [13], a framework for designing textual DSLs. MontiCore uses context free grammars as input artifacts to generate lexer, parser, prettyprinter, abstract syntax tree (AST) classes, runtime components and editor support for DSLs. Furthermore, it enables context condition checking [10], symbol table implementations [14], code generation and mechanisms for integrating handwritten code [3]. Context conditions are syntactically checkable rules that further constrain the set of valid instances of a language. In this paper, we specify context conditions in a mathematical way to provide conformance rules between models of the derived languages. For generating code we use Freemarker [15] as a template language for a model to text transformation.

MontiCore combines concrete and abstract syntax within a single grammar file and uses a syntax similar to EBNF. Within such a grammar that starts with the keyword `grammar` followed by its name, productions consisting of a nonterminal and a right-hand side (RHS) that specifies attributes and compositions within the AST. As with

EBNF, the right-hand side may contain terminal or nonterminal symbols. Also, alternatives ("|"), optionality ("?") and multiple cardinalities ("+", "*") are supported. Apart from that MontiCore offers several extensions over EBNF, such as the use of interfaces and extending existing grammars, denoted by the keywords `implements` and `extends` [12], [13]. Similar to the corresponding polymorphic concepts, interfaces can be implemented by other productions and may be used as nonterminals. Such interfaces may even be implemented in sublanguages in order to reuse the production of the parent language. Basically, this mechanism is an extended form of alternatives where the single alternatives are not known beforehand. In general, extending existing grammars increases the modularity and reuse since the extending grammar can directly use all nonterminals from the extended grammar.

For integrating different languages, MontiCore offers three distinct mechanisms: language aggregation, language embedding and language inheritance. Language aggregation allows to combine two different languages in separate artifacts by referencing elements and establishing a knowledge relationship between both heterogeneous artifacts. Language embedding allows the combination of two different languages into a single artifact by defining external nonterminals that are filled with nonterminals of the embedded language. This is especially useful for embedding different paradigms, such as embedding behaviour modeling capability into a structural language. Finally, language inheritance offers the possibility to create sublanguages and to reuse the productions of the parent languages via the grammar extension mechanism. This mechanism also supports multi inheritance. Our approach makes use of language inheritance and language aggregation. A more detailed discussion on these three mechanisms can be found in [14], [16], [17].

## IV. Methodology

In this section the process to derive the tagging languages for a given DSL is introduced. Derivation means that following a set of given derivation rules leads to the systematic development of new DSLs. Before the specific derivation rules for both languages are presented, Figure 2 gives an overview about the involved languages, corresponding models and the relations between them. Afterwards, languages containing the common production for the tagging languages are introduced. The process for deriving domain specific tagging languages is shown. It is based on MontiCore as the existing modeling language, and both derived tagging languages are defined as MontiCore grammars. To ensure the consistency between all models, context conditions are presented. The language application section describes involved roles and activities needed to create tagging models.

The different languages involved are shown in the lower part of the figure. The MontiCore Grammar (MCG) flag on the right indicates that the languages are defined as MontiCore grammars. For an existing source language $L_G$ the tag language $L_G^{Tag}$ and the tagschema language $L_G^{Schema}$ can be derived following the process presented in Sections IV-C and IV-D. $L_G^{Tag}$ builds on both $L_{COMMON}^{Tag}$ and $L_G$, whereas $L_G^{Schema}$ only builds on $L_{COMMON}^{Schema}$ but depends on $L_G$. The builds on relation is implemented using language inheritance. That means that $L_G^{Schema}$ includes concepts of the abstract syntax of $L_G$ as concrete syntax (shown in detail in Section IV-D). In contrast, the depends on relation means that $L_G$ is used within the derivation process without having an inheritance relation between $L_G^{Schema}$ and $L_G$. Instead the productions of $L_G^{Schema}$ depend on specific parts of the concrete syntax of $L_G$. The predefined common languages $L_{COMMON}^{Tag}$ and $L_{COMMON}^{Schema}$ encapsulate productions to be reused by the tagging languages. Since many concepts can be reused within sublanguages, this eases the effort for creating tag languages as the number of derivation rules can be reduced substancially. The two common languages provide interfaces that have to be implemented by the sublanguages. Within the common language, the interface type can be used and thus, most productions are already defined in the common tagging languages. These common languages are supposed to stay fixed for the derived DSLs.

In the upper part of the figure, models of the different languages are shown. The MontiCore model (MCM) flag on the upper right corner indicates that these models can be processed by languages of the MontiCore language workbench. As shown in the example of SCs in Section II, the models $M_{L_G^{Tag}}$ of $L_G^{Tag}$ reference $M_{L_G}$ and its elements. Furthermore, the models $M_{L_G^{Schema}}$ of $L_G^{Schema}$ contain references to elements of the language $L_G$, which are the types of elements of $M_{L_G}$. Moreover, references between the models $M_{L_G^{Tag}}$ and the corresponding $M_{L_G^{Schema}}$, as well as their elements, exist. References between models are established using the language aggregation mechanism of MontiCore.

### A. The $L_{COMMON}^{Tag}$ Language

Listing 4 shows the MontiCore grammar of $L_{COMMON}^{Tag}$. The $L_{COMMON}^{Tag}$ language serves as a basis of $L_G^{Tag}$. It defines the overall structure of all tag models. The `Tag-Model` production marks the beginning of a tag model. First a comma separated list of `QualifiedName` can be specified in order to reference corresponding tagschemas, explained in Section IV-B. A tag model then starts with the keyword `tags`, followed by a `QualifiedName` referencing the actual target model whose information is tagged.

The grammar provides two major concepts and the interface `ModelElementIdentifier` that has to be implemented in sublanguages. The first concept is the concept of `Contexts` that enable navigating into elements, via the `within` construct. As shown in Listing 4, the `Context` production uses the `ModelElementIdentifier` interface to address elements. Other contexts may be

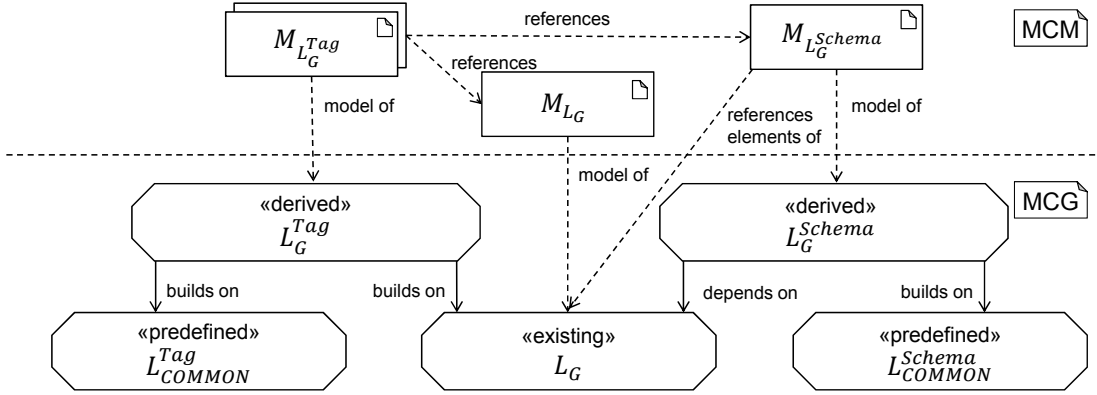

Fig. 2: Overview of involved languages, corresponding models and the relations between them.

```
1  grammar Tags extends Common{
2
3    TagModel =
4      "conforms" "to"
5        QualifiedName ("," QualifiedName)*";"
6      "tags" Name "for" targetModel:QualifiedName
7      "{" (contexts:Context | tags:TargetElement)* "}"
8    ;
9
10   Context = "within" ModelElementIdentifier "{"
11     (contexts:Context | tags:TargetElement)*
12     "}"
13   ;
14
15   interface ModelElementIdentifier;
16
17   DefaultIdent implements ModelElementIdentifier =
18     QualifiedName
19   ;
20
21   interface Tag;
22
23   TargetElement =
24     "tag" ModelElementIdentifier
25     ("," ModelElementIdentifier)*
26     "with" Tag ("," Tag)* ";"
27   ;
28
29   SimpleTag implements Tag = Name;
30
31   ValuedTag implements Tag = Name "=" String;
32
33   ComplexTag implements Tag = Name
34     "{" (Tag ("," Tag)* ";")? "}"
35   ;
36 }
```

Listing 4: The $L_{COMMON}^{Tag}$ MontiCore grammar.

nested within in order to navigate further into the hierarchy or tags for specific elements. The second concept is the `TargetElement` production, defining the tagging of a specific element. It starts with the keyword `tag` followed by a comma separated list of `ModelElementIdentifier`. After that a `with` follows and then the actual tagged information represented by the interface `Tag`. The list of `ModelElementIdentifier` enables simultaneously tagging of multiple elements. The interface itself is implemented in sublanguages for the purpose of uniquely identifying concrete elements of $M_{L_G}$. The derivation and implementation of the sublanguages is shown in Sections IV-C and V.

The interface `Tag` is directly implemented within $L_{COMMON}^{Tag}$ by three different productions. The `SimpleTag` consists only of its name in order to flag an element of the target model with additional information. The `ValuedTag` tag consists of a name and an additional value. This value is always represented as a String in $L_{COMMON}^{Tag}$ but, in combination with the corresponding tagschema, it can be another data type (such as numbers) as well. This is explained in Section IV-B

The `ComplexTag` tag allows to express complex information with multiple subtags. The name of the tag is followed by the nested tag block (enclosed in curly brackets), which consists of a comma separated list of subtags.

### B. The $L_{COMMON}^{Schema}$ Language

Beneath the common language $L_{COMMON}^{Tag}$ we also created a common language to be used by language specific tagschema languages, called $L_{COMMON}^{Schema}$. The language $L_{COMMON}^{Schema}$, shown in Listing 5, follows the same notion as $L_{COMMON}^{Tag}$. Within the common language again, all necessary structuring productions and an interface that has to be implemented by the sublanguages is already defined. The tagschema language starts with the keyword `tagschema` followed by a name. This name is the identifier that has to be used in the `conforms to` element of the tag model, shown in the previous section, in order to reference the tagschema model.

Within the tagschema model, multiple `TagTypes` can be specified. Each tag type may be private, meaning that this type can only be used within other tag types and not as a first level type. In general, each `TagType` starts with the keyword `tagtype`, has a *Name*, that is unique within a schema, and has a `ScopeIdentifier`. The scope references nonterminals of $L_G$. Since the scope is language specific, it is represented by an interface in $L_{COMMON}^{Schema}$ and has to be implemented in $L_G^{Schema}$, where the different scopes are known.

The interface `TagType` is implemented by four specific kinds of types: `SimpleTagType`, `ValuedTagType`, `EnumeratedTagType`, and `ComplexTagType`. A

```
1  grammar TagSchema extends Common {
2
3    TagSchema = "tagschema" Name
4      "{"
5          TagType*
6      "}"
7    ;
8
9    interface TagType;
10
11   interface ScopeIdentifier;
12
13   Scope = "for"
14     (ScopeIdentifier ("," ScopeIdentifier)* | "*")
15   ;
16
17   SimpleTagType implements TagType =
18     ["private"]? "tagtype" Name Scope? ";"
19   ;
20
21   EnumeratedTagType implements TagType =
22     ["private"]? "tagtype" Name ":"
23     "[" String ("|" String)* "]" Scope? ";"
24   ;
25
26   ValuedTagType implements TagType =
27     ["private"]? "tagtype" Name ":"
28     ("int"|"String"|"Boolean") Scope? ";"
29   ;
30
31   ComplexTagType implements TagType =
32     ["private"]? "tagtype" Scope?
33     "{" Reference ("," Reference)* ";" "}"
34   ;
35
36   Reference = Name ":" ReferenceTyp ("?"|"+"|"*")? ;
37
38   ReferenceTyp = ("int"|"String"|"Boolean"|Name);
39 }
```

Listing 5: The $L_{COMMON}^{Schema}$ MontiCore grammar.

`SimpleTagType` defines a tag type which solely consists of a key as a simple, static flag. In $M_{L_G^{Tag}}$ such tag types are expressed as a `SimpleTag`. A `ValuedTagType` contains a value which can be of a defined type, such as `int`, `String` or `Boolean`. In $M_{L_G^{Tag}}$ such tag types are expressed as a `ValuedTag` with a String value. The type checking is ensured via context conditions, explained in Section IV-E. An `EnumeratedTagType` defines a set of values that compose the domain of that tag type. An instance of an enumerated tag type, modeled in $M_{L_G^{Tag}}$ is also expressed as a `ValuedTag` and must additionally declare a value that exists in the defined domain. Thus, the enumerated tag type is restricted in its values, whereas the valued tag type is more expressive and only restricted by its native type. A `ComplexTagType` has the most expressive power as it can define an arbitrary number of subtag types. Such subtag types are again directly defined or referenced tag types, nested within the complex tag type. A reference has a unique name within the complex tag type, and a certain type. The type may either be a primitive data type, analogously to the `ValuedTagType`, or a reference to another tag type already defined in the schema. Hence it is possible to nest information of various tag types into one complex tag and there is no limitation on the depth of the hierarchy.

Furthermore, each subtag may have a cardinality which can be one of three kinds: optional ("?"), arbitrary many ("*") or at least one ("+"). If the cardinality is left out, the subtag is required. In $M_{L_G^{Tag}}$ such tag types are modeled as `ComplexTag` with the subtags nested in the curly brackets.

### C. $L_G^{Tag}$ Derivation Process

To derive a $L_G^{Tag}$ for a given source language $L_G$, we utilize the inheritance mechanism of MontiCore. In this way, production rules of $L_{COMMON}^{Tag}$ and $L_G$ can be reused and additional rules can be specified.

**Addressing Elements:** As it is possible to tag elements that are nested within others (e.g. nested states), there are two different ways to address those elements. The first possibility is to use a dot-separated syntax (e.g. `Active.Call`) but this technique may get tedious when trying to tag multiple elements or if the elements are nested deeper. The second possibility is the use of *contexts*. A context consists of the keyword `within` followed by an identifier and its body (enclosed in curly brackets, cf. Listing 2, l.8). XPath like path definitions have been eliminated, since they are too unstable against changes in the models. XQuery is more for tagging sets of elements, where we use transformations. The identifier refers to a target element, which is used as a context to resolve all other elements that are defined within the body of the context.

In general, the resulting tag language should be able to address all elements specified by models of $L_G$. The type of those elements is given by the nonterminals $N \in L_G$ in the source language grammar. To facilitate the derivation, the $L_{COMMON}^{Tag}$ offers the `ModelElementIdentifier` (cf. Listing 4, l.15) interface that has to be implemented by the productions of the sublanguage $L_G^{Tag}$. By implementing this interface, elements of $M_{L_G}$ become identifiable and can be referenced by $M_{L_G^{Tag}}$. To address those elements, one of the following rules have to be applied:

**IV-C.1.** *Let $N \in L_G$ be the set of nonterminals of $L_G$. For every $n \in N$ that can be identified by a qualified name, the `DefaultIdent` rule of $L_{COMMON}^{Tag}$ can be reused (cf. Listing 4, l.17).*

**IV-C.2.** *For every other nonterminal $n \in N$, a manual identifier has to be chosen, that uniquely identifies the element. This is dependent on the semantics and the use of the language. A general possibility to ensure a correct identification rule is to reuse the concrete syntax of the tagged element in the tagging language. Therefore, a new rule $I_n$ is added to $L_G^{Tag}$ where the nonterminal of $L_G$ enclosed in brackets is used to address corresponding elements:*

$$I_n \text{ implements } ModelElementIdentifier = "[" n "]"$$

Nevertheless, reusing the concrete syntax of the tagged element typically leads to a poor concrete syntax of the tagging language. If there is a different identifier suitable, this should be taken. This can enhance the usability of the tagging language. The rule must only ensure that corresponding elements can be addressed uniquely.

### D. $L_G^{Schema}$ Derivation Process

To derive $L_G^{Schema}$ for a given source language $L_G$, the inheritance mechanism of MontiCore is utilized again. In this way, production rules of $L_{COMMON}^{Schema}$ can be reused and additional rules can be specified. In contrast to $L_G^{Tag}$ and $L_G$, there is no inheritance relation between $L_G^{Schema}$ and $L_G$. Nevertheless, $L_G^{Schema}$ depends on $L_G$ as parts of the concrete syntax are reused for the definition of $L_G^{Schema}$ (instead of referencing existing rules).

**Addressing Element Types:** In addition to the rules of $L_{COMMON}^{Schema}$, derivation rules to address element types of $M_{L_G}$ must be added to $L_G^{Schema}$. This is necessary as the tagschema language is used to define which tags can be added to which type of elements. To facilitate the derivation the $L_{COMMON}^{Schema}$ offers the ScopeIdentifier (cf. Listing 5, l.11) interface that has to be implemented by the productions of the sublanguage $L_G^{Schema}$. By implementing this interface, elements of $L_G$ become identifiable and can be referenced by $M_{L_G^{Schema}}$. The fact that each type of an element is defined by a nonterminal of the corresponding MontiCore grammar is used to define the derivation rule for $L_G^{Schema}$:

**IV-D.1.** *Let $N \in L_G$ be the set of nonterminals of $L_G$. For every $n \in N$ a new rule $SI_n$ is added to $L_G^{Schema}$:*

$$SI_n \; implements \; ScopeIdentifier = "n";$$

**Addressing Nested Element Types:** We need to consider that nonterminals of $L_G$ might be used more than once on the RHS of a production, e.g. "source:Name '->' target:Name", taken from the SC grammar, where the nonterminal Name is used twice. Within MontiCore, such occurrences of nonterminals are distinguished by identifiers preceding the nonterminal. As we need to address each occurrence in $L_G^{Schema}$, we add additional production rules implementing the ScopeIdentifier interface, which utilize the preceding identifiers of nonterminals of $L_G$.

**IV-D.2.** *Let $PI_{st}$ be the set of preceding identifiers that occur more than once on the RHS of a single production $st \in LG$. For each preceding identifier $pi \in PI_st$ for all st, a specific production rule $SI_{pi}$ is added to $L_G^{Schema}$ in order to address the different occurrences in the scope st defining nonterminal n:*

$$SI_{pi} \; implements \; ScopeIdentifier = "n\_pi";$$

Note that all identifiers are unique within a single production even if they are used to distinguish different nonterminals on the RHS. If the identifer is unique in the scope of all $st \in L_G$, the prefix $n\_$ can be omitted.

### E. Context Conditions

Figure 2 already gave an insight into the relations between models of the different modeling languages. These relations where bounded to the technical relations references and references elements of. In fact, the relations between the models are more complex and the conformance of models among each other have to be validated by additional rules in contrast to the consistency between all involved languages where no additional context conditions are required, since they are ensured by construction through the derivation rules. The following context conditions are used to consistency check the tag model as well as the tag schema model.

Before introducing these context conditions, some variables are defined: Let $st \in M_{L_G^{Tag}}$ be a tag defining statement of $M_{L_G^{Tag}}$, $MR_{st}$ the set of model element references of $st$, $TR_{st}$ the set of tag type references of $st$, $tr.value : tr \in TR_{st}$ the value, either native, enum or complex, assigned to a tag, $Dom(tr.value)$ the type, either specific native, enum or complex, of the value, $tt \in M_{L_G^{Schema}}$ a tag type defining statement of $M_{L_G^{Schema}}$, $tt.name$ the name of the defined tag type, $tt.domain$ the type of the value which can be assigned to this tag type, and $MTR_{tt}$ the set of model element type references of $tt$ defined in $L_G$.

**Tagged elements:** A relation between $M_{L_G^{Tag}}$ and $M_{L_G}$ is explicitly stated within the $M_{L_G^{Tag}}$ (cf. Listing 2, l.4). The correctness of this relation is given if each tagged element of $M_{L_G^{Tag}}$ exists in $M_{L_G}$.

**IV-E.1.** *For each $st \in M_{L_G^{Tag}}$, $mr \in MR_{st}$:*

$$mr \in M_{L_G}$$

**Conforms to:** The conforms to relation between a $M_{L_G^{Tag}}$ model and a corresponding $M_{L_G^{Schema}}$ model is also stated explicitly (cf. Listing 2, l.2). The correctness of this relation is given if for every tag statement in the $M_{L_G^{Tag}}$ a corresponding tag type is defined in $M_{L_G^{Schema}}$.

**IV-E.2.** *Each tag type of $M_{L_G^{Schema}}$ has a unique name which allows the unique mapping between used tag types within statements of $M_{L_G^{Tag}}$ and the tag type definition statements of $M_{L_G^{Schema}}$. For $tt_1, tt_2 \in M_{L_G^{Schema}}$:*

$$tt_1.name = tt_2.name \Rightarrow tt_1 = tt_2$$

If multiple schemas are referenced by $M_{L_G^{Tag}}$, the rule has to be extended to ensure unique names for all included tag types of all referenced tagschemas.

**IV-E.3.** *For each tag type reference of each statement of $M_{L_G^{Tag}}$, a corresponding tag type exists (1). Furthermore, the model element types of tagged elements fit to the types defined in $M_{L_G^{Schema}}$ (2) and the types of tag values fit to* the types also defined in $M_{L_G^{Schema}}$ (3). Thus, for each $st \in M_{L_G^{Tag}}$, $tr \in TR_{st} : \exists\, tt \in M_{L_G^{Schema}}$:

$$tt = tr \tag{1}$$
$$\wedge \quad \{mr.type \mid mr \in MR_{st}\} \subseteq MTR_{tt} \tag{2}$$
$$\wedge \quad Dom(tr.value) = tt.domain \tag{3}$$

Note that the comparison of tag value types (3) must support native types, enum types and complex types.

### F. Language Application in MDE

So far, a methodology for deriving the tagging languages for a given source language has been presented. In this paragraph the application of both languages in a MDE environment is examined. For simplicity, it is assumed that a single generator is involved in the MDE process which takes $M_{L_G}$ models together with corresponding $M_{L_G^{Tag}}$ models as input.

Before the generator can be used, the generator has to be developed by the generator developer. Beside the usual generator development steps, the generator developer models $M_{L_G^{Schema}}$ and provides the model as part of the generator. The generator is developed in a way that multiple $M_{L_G^{Tag}}$ conforming to the $M_{L_G^{Schema}}$ can be processed. The generator user is the role that creates the inputs for the generator, which are the domain model $M_{L_G}$ and in addition a model $M_{L_G^{Tag}}$ for the additional, e.g. domain specific, information. Of course this role can be taken by more than one person, most likely an expert for each domain, such as a domain expert and a database designer who tags the domain model with database specific configuration information.

In a more complex scenario, several $M_{L_G^{Schema}}$ can be provided by the generator developer. This implies that the generator must be able to process multiple $M_{L_G^{Tag}}$ conforming to any of the $M_{L_G^{Schema}}$. This enables a clean model-based specification of all necessary additional information not in the scope of the domain. Note that in this complex scenario, the languages $L_G^{Schema}$ and $L_G^{Tag}$ must not be adapted concerning the simpler scenario.

## V. CASE STUDY

In order to validate our approach, we applied it to three different languages: SCs, CDs and directly to MontiCore grammars. The case study is used to evaluate if the method itself is applicable to a set of languages and to get an impression on the benefits for the different roles. It does not provide a solid empirical study. Apart from the sole applicability, we check if creating the tagging language can be done with small effort. Additionally, we check if we can check the tag models against the schema, i.e. we are able to define constraints on the target element types and on the structure of tags in general. Furthermore, the possibility of creating multiple models was explored.

Figure 3 shows the application to an excerpt of the SC grammar. The SC grammar also inherits some nonterminals from the Common language, the $L_{COMMON}^{Tag}$ and $L_{COMMON}^{Schema}$ languages also inherited from. Furthermore, the SC grammar consists of the nonterminals SCDefinition, Transition, TransitionBody, State and Invariant. The Element interface and the TransitionBody are omitted within the derivation process. For the other nonterminals, the derivation process for deriving $L_G^{Tag}$ is shown in Figure 3 and the necessary steps are applied. Implicitly, every SCDefinition and every State can be identified by a unique name. Thus, the default production for ModelElementIdentifier defined in $L_{COMMON}^{Tag}$, following step IV-C.1, is used. For the remaining two nonterminals, the derivation rule IV-C.2 that reuses the abstract syntax of $L_G$ is applied, as shown in Figure 3. The two nested source and target variables are also directly identifiable by a name and therefore can be tagged via the transition identifier and the respective source or target name. An instance of this derived language was already shown in Listing 2. The names of the SC and the states have been used for identifying the elements. The presented model can be parsed with the parser generated by MontiCore out of the derived language. The complete SC language consists of 16 different productions and six nested nonterminals that have to be distinguished by a variable name. Six nonterminals are not uniquely identifiable by a name. Thus, the derived tag language includes six productions to specify these identifiers. For Classdiagrams the situation is similar: The grammar contains 17 productions with 10 nested nonterminals that have to be distinguished by a variable name and the derived tag language contains five productions. The grammar for MontiCore grammars presents itself a bit different since we chose to tag only certain elements, i.e. the grammar itself and productions. Thus, the corresponding tag language contains only two productions.

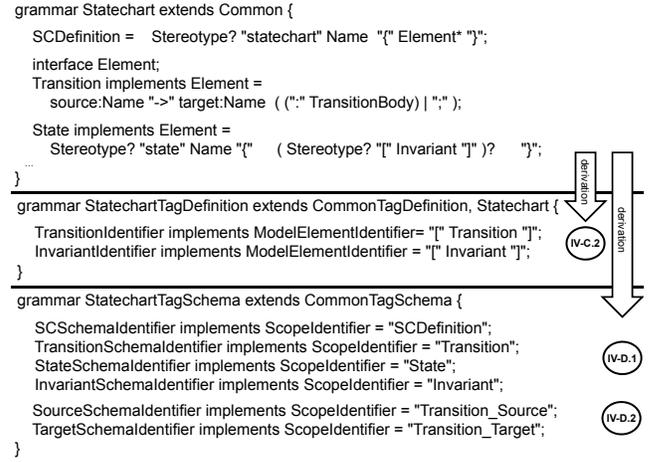

Fig. 3: Application of the tagging languages derivation rules for the SC grammar.

The derivation of $L_G^{Schema}$ is shown in Figure 3. We again apply the derivation rules and use the name of the nonterminal as an identifier. This is done for the SC itself, states, transitions and invariants in order to address these elements in a tagschema. Furthermore, the nested source and target of a transition are also included since there may be tags that are only allowed at these elements. An instance of this derived language was already shown in Listing 3. The resulting schema language for the SC language consists of 22 productions implementing the `ScopeIdentifier`. For the CDs and the MontiCore grammar language there are 27 and two productions, respectively.

While these three applications are certainly not enough for a solid empirical validation, we have high confidence that developers of tagging languages, generator developers and domain experts, i.e. product developers, will have considerable benefits of the approach especially in larger, multi-tier projects, where configuration and other technical information are complex. We were able to show that it is possible to apply our method with small effort, which should help language developers in creating these languages. None of the languages needed a lot of additional productions. We were also able to automatically take schemas into account for the languages mentioned supporting the generator developer. Seperation of concerns was also possible by creating multiple models with different information, supporting domain experts.

## VI. Related Work

Similar to the approach of [18], the $L_G^{Tag}$ language can be automatically derived from the target grammar G by using the concrete syntax of the target element type as its identifier. In our work we recommend to manually choose a readable identifier and provide a default derivation rule for systematic derivation. In [18] the default rules are generated and can be overridden afterwards by a language extension. An alternative approach where variability is added to DSLs is presented in [19].

Related is the ProMoBox approach [20], that generates five different sublanguages out of a given metamodel of a DSL. These languages are used to specify certain aspects of the system, such as design, input, output, runtime and properties. This approach differs in the fact that the tags in our case are used for a different purpose, such as enriching the model with arbitrary information. Other approaches aim at reducing the complexity of a model using multilevel modelling approaches [7], [21].

In [22] three primary methods for defining a DSL are presented. Two of those extend or refine an existing modeling language. The source language is the UML [23] whose extension techniques can be compared to our approach as we build upon a source language as well. The first mechanism is to add stereotypes to elements of the source language which enables specialized DSL tools to display the marked elements in a different way. This is similar to our approach which tags special elements of the source language. However, in our approach models of the source language do not need to be changed. Thus, we avoid model pollution by separating the information in several artifacts.

The second described technique is UML profiles [23] where parts or the whole UML metamodel can be reused to define DSLs or domain-specific viewpoints. This approach is limited to UML while our approach can be applied to DSLs. Moreover, in [22] a systematic method to define such UML Profiles is presented, which is comparable to our approach but more abstract. Furthermore, [24] presents an automated approach to extract UML profiles from Java annotation libraries. UML profiles have been extended to EMF profiles in [25]. We will investigate how to transfer the results to our approach in order to automatically derive tagschema models. Most work extends the UML by suited stereotypes or profiles, such as [26] for embedded real-time systems, [27] for Web applications, [28] as a profile for AADL applications, and even for software product lines [29]. While there are many profiles available for all kinds of concerns, these are all restricted to the UML and not to DSLs in general.

Furthermore, a lot work has been done in the area of transformation languages. An overview of different model transformations is given in [30] and [31]. In this area, some work can be found that focuses on the derivation of domain specific transformation languages, following a systematic approach [32], [33] or taking the concrete syntax of the source language into account [33], [34], [35]. Such transformations can be used to transform technology agnostic models to technology aware models, which is supported by several transformation tools [36], [37], [38], [39].

## VII. Conclusion

In the presented work a methodology for deriving a language specific pair of tagging languages has been introduced. Built upon two common parent languages, the effort for creating the language specific parts could be minimized. Beside the systematic stepwise derivation process, necessary context conditions for ensuring consistency and validity of the models have been presented. Moreover, advices for the application of the tagging languages have been given and a case study ensuring the feasibility of the approach has been presented. For future work we aim at implementing our methodology in an automated language generation process, that makes use of reasonable defaults and is able to generate both languages directly from a given grammar. Furthermore, the generation of corresponding context conditions is planned which will allow the automatic validation of corresponding tag and tagschema models. Additionally we aim at exploring conflict resolution strategies for tags and the possiblity to tag tags themselves. As additional tool support, the generation of a tagschema model editor from source grammars and the generation of tag model editors from corresponding tagschema models is planned. Moreover, back-end infrastructure for the generator could be generated to make processing of tag models easy.

Here, the automatic generation of a class hierarchy from tagschema models would allow to represent information of tag files conforming to a tagschema model by objects of the generated class hierarchy. This would enable the generator developer to access the additional information from tagschema models in a well typed manner.